# Spectral Characterization of Polyvinylidene Difluoride Sensor Designs for Acoustic Spectroscopy

Ishan V. Ramaiah[1], Yuri A. Pishchalnikov[1], and William M. Behnke-Parks[1]

[1]*Applaud Medical, 953 Indiana St., 94107, San Francisco, California, USA*

Polyvinylidene difluoride (PVDF) sensors are of interest in ultrasonic applications due to their high sensitivity and efficient acoustic coupling to the environment. PVDF sensors for scientific applications typically possess a flat frequency response to faithfully reproduce spectral components of an unknown source. However, in industrial sensing applications higher responsivity is desirable even at the cost of decreased flatness in spectral responsivity of the sensor. To quantify the trade-off between increased intrinsic responsivity and non-linearity of the spectral response we examine four PVDF sensor variants for spectral responsivity as a function of thickness and find that increased responsivity is correlated with a strong non-linear frequency response.

**I. Introduction**

PVDF is a widely used component in sensors for applications such as motion detectors, robotics, and acoustic spectroscopy [1-4]. Once polarized, PVDF are ideal sensors for acoustic detection due to their high dielectric sensitivity and good coupling efficiency with the environment due to the low acoustic impedance (~4 MRayl).

A typical broadband PVDF sensor is approximately 10 to 20 µm thick uniaxially poled PVDF membrane between opposing electrodes with a thickness-mode resonant frequency of approximately 50 to 100 MHz allowing for a broad flat spectral response below 10 MHz [5][6]. The rational for such a design is to minimize deconvolution needed when comparing spectra across a broad frequency range. Since the responsivity is a function of the electrical impedance and mechanical response, increasing the membrane thickness is expected to increase the mechanical displacement and electrical sensitivity for a given aperture geometry. However, there may be a concomitant change in the spectral response, and characterizing this change is the objective of this study.

In this study, we investigated the responsivity of four PVDF sensor thickness variants (50 µm, 80 µm, 110 µm, 220 µm). The responsivity of an acoustic sensor can be subdivided into the following contributions: the sensor's intrinsic electro-mechanical responsivity, the interface coupling efficiency, and coupling efficiency with the environment based on the aperture geometry. To measure the intrinsic membrane responsivity of each sensor in isolation from other contributions, we implement intermediate-field acoustic measurements with PVDF membranes mounted on acoustic absorbers combined with temporal windowing to measure the steady state responsivity. To explore the sensor's spectra response, each sensor was insonated with each of four different frequencies typically used in therapeutic applications (0.25 MHz, 0.5 MHz, 1.0 MHz, and 2.0 MHz) from four different transducers. The acoustic output power of each transducer was measured separately with a high-precision calibrated radiation force balance to normalize the sensor response at a given frequency.

## II. Methods

The transducers used were the PA 713 ($f_0$ = 250 kHz), PA 711 ($f_0$ = 1 MHz), PA 706 ($f_0$ = 2 MHz) from Precision Acoustics (Dorchester, U.K.), and a single channel variant of the XDR089E ($f_0$ =500 kHz) from Sonic Concepts (Bothell, WA, U.S.A.). The acoustic power of each transducer was measured using a Radiation Force Balance (Precision Acoustics)[7] calibrated at the National Physics Laboratory (Teddington, U.K.). An opening with similar dimensions and area within ~10% to the PVDF sensor was cut into two layers of acoustic absorbing material (F28, Precision Acoustics) suspended above the target at 4 cm from the face of the transducer. The power of each transducer was then scaled for duty cycle for normalization purposes. The power output of each transducer is listed in Table 1.

*Table 1: Acoustic power measurements from transducers measured using the acoustic radiation force method (n=10, mean ± s.e.m.).*

| Condition | PA 713 | XDR089 1-chan. | PA 711 | PA 706 |
|---|---|---|---|---|
| Total Power Output (mW) | 3350 ± 10.08 | 1030 ± 1.98 | 1370 ± 5.46 | 248 ± 0.51 |
| Power Output on PVDF (mW) | 1180 ± 2.47 | 590 ± 2.36 | 183 ± 0.27 | 52.0 ± 0.27 |

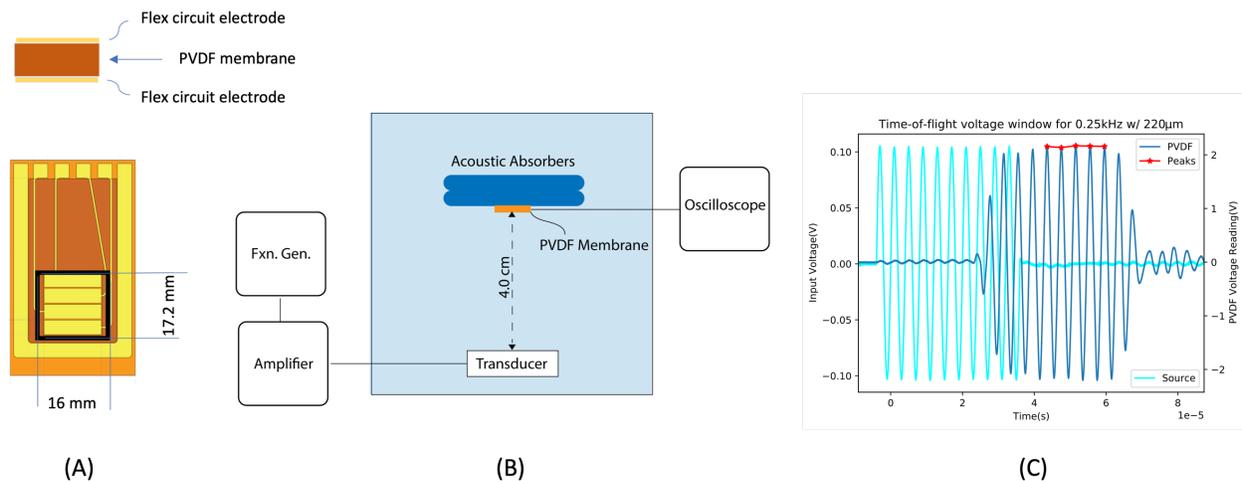

*Figure 1: Experimental set-up. (A) The PVDF sensor design comprises a flex circuit electrode pattern flanking a poled PVDF membrane of variable thickness. (B) The intermediate-field measurements were conducted at 4.0 cm with each sensor variant mounted to two acoustic absorbers. (C) The windowing algorithm consisting of selecting peaks 4-9 ordered temporally from the set of ten maximum peak voltages (red). This algorithm accounts for time-of-flight required to reach steady-state.*

The PVDF variants were custom build (Precision Acoustics) flex circuits on a PVDF membrane (see Figure 1). The PVDF membrane thickness was 50 μm, 80 μm, 110 μm, or 220 μm. The 220 μm PVDF membrane was constructed by laminating two 110 μm membranes together. PVDF sensors of the same design were used to verify reproducibility of results.

In all experiments, the ground plane was adhered to the acoustic absorber. The acoustic absorber (F28, Precision Acoustics)[8] was stacked in a double layer mounted to an aluminum plate to further suppress acoustic reflections. The double layer resulted in a total thickness equivalent to approximately 40 mm of F28. The Fractional Power Dissipation (FPD) for this double layer can be calculated from:

$$FPD = 1 - (P_t/P_i)^2 - (P_r/P_i)^2$$

$$FPD = 1 - 10^{2(IL/-20)} - 10^{2(ER/-20)}$$

Where IL is insertion loss, and ER is echo reduction. IL is defined as:

$$IL = -20\log_{10}(P_t/P_i)$$

The worst-case IL value for F28 material between 0.25 MHz and 2 MHz is 10 dB per cm for a total of 40 dB. The Echo Reduction (ER) can be calculated from:

$$ER = -20\log_{10}(P_t/P_i)$$

Where $P_r$ is the amplitude of the acoustic pressure reflected from a sample and $P_i$ is the amplitude of the acoustic pressure incident upon it. The worst-case ER value is 25 dB per interface for four surfaces for echo reduction, ER = 100 dB. Calculating FPD from these IL and ER values yields a worst-case FPD of 99.99% for the frequency range (0.25 – 2 MHz) used in this study. In all cases, the time-of-flight for the entire pulse was sufficiently brief relative to the dimensions of the water tank as to avoid overlap with additional reflections, including secondary reflections from the face of the transducer equal to or greater than a factor of three times the time-of-flight of the initial, direct insonating pulse.

The responsivity was measured under steady-state insonation by windowing the acoustic signal received in an intermediate-field regime of 4.0 cm transducer-to-sensor distance (see Figure 1). A pulse of 10 cycles at the corresponding frequency were sent to the corresponding transducer via a frequency generator (Tektronics, AFG2021) and power amplifier (ENI AP 400). Steady-state was achieved after a pulse duration greater than 3.8 microseconds based on a maximum 44 mm diameter aperture and differential time-of-flight. Data was acquired using an oscilloscope (Teledyne LeCroy WaveSurfer 4104HD) set to 1 MΩ transimpedance resistance. The acoustic window was defined as pulses 4 through 9 ordered by arrival time selected from the 10 acoustic peaks in a pulse. This ensured the electrical signal and acoustic signal did not overlap and allowed sufficient time of flight to achieve steady state.

## III. Results and Discussion

The responsivity of each PVDF variant was characterized for given frequencies and plotted as a function of thickness. PVDF sensor responsivity displayed a general trend towards increasing responsivity with increasing PVDF membrane thickness (Figure 2). This is consistent with the expected increased thickness resulting in a larger pressure-driven displacement, increasing the change in capacitance and therefore the electrical signal produced.

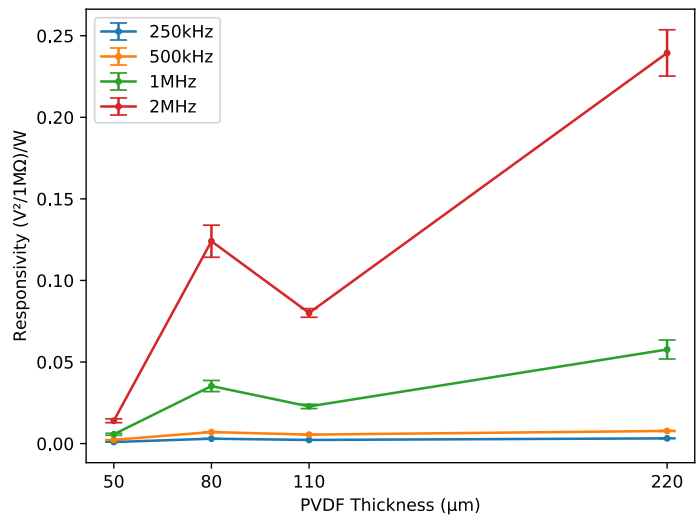

*Figure 2: PVDF sensor responsivity versus sensor thickness for various frequencies (n = 4, mean ± s.e.m.). Lines between points are guides to the eye.*

Interestingly, the responsivity was not monotonically increasing, as the 110 µm variant showed a slight decreased responsivity relative to the 80 µm variant. This may be due to inefficiencies in the poling of 110 µm PVDF membrane due to the higher voltage required producing unproductive sparking events. Note that the PVDF 220 µm variant is constructed by adhering two 110 µm PVDF membranes together and therefore continues the trend towards increased responsivity with increasing PVDF membrane thickness.

Plotting PVDF variant responsivity versus insonation frequency illustrates a strongly non-linear dependence of responsivity with frequency for increased PVDF membrane thickness. Fitting the data to a power law of the form $\xi = Af^{\alpha}$, where $\xi$ is responsivity, $f$ is frequency and $A$ and $\alpha$ are free parameters illustrates this non-linear dependence (Figure 3) with the resulting fit parameters displayed in Table 2.

The increasing power $\alpha$ with increasing PVDF membrane thickness suggests spectral distortion increases with increasing PVDF membrane thickness. We hypothesize that the resonant frequency drops as the variant thickness is increased and that the spectral responsivity shown in Figure 3 represents the shoulder of this resonant peak. Modeling the resonant frequency as a purely mechanical waveguide, the acoustic impedance due to waveguide effects will drop to zero and the resonant frequency when the thickness ($\delta$) is half of an acoustic wavelength ($\lambda$):

$$\delta = \frac{\lambda}{2} = \frac{c}{2f_0}$$

where c is the speed of sound in PVDF (c = 2200 m/s). The calculated resonant frequency $f_0$ for the $\delta$ = 220 μm variant is 5 MHz, consistent with the strong non-linear dependence of responsivity on frequency reflecting the shoulder of a resonant peak.

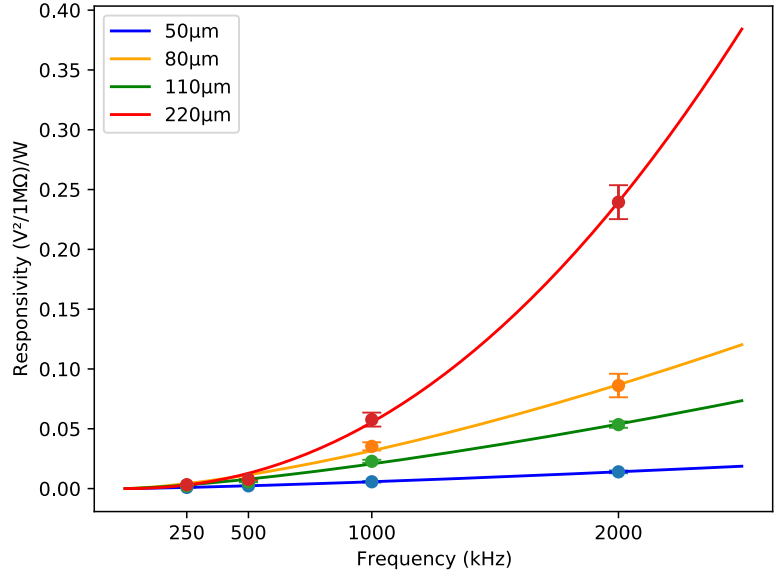

Figure 3: PVDF sensor responsivity versus insonation frequency for various PVDF thickness (n = 4, mean ± s.e.m.) fit to a power law of the form $\xi = Af^{\alpha}$ where $\xi$ is the responsivity, f is the frequency and A and $\alpha$ are free parameters.

Table 2: Fit parameters from Figure 3 using a power law of the form $\xi = Af^{\alpha}$ where $\xi$ is the responsivity, f is the frequency and A and $\alpha$ are free parameters, and $R^2$ is the goodness of fit.

| PVDF thickness | A | $\alpha$ | $R^2$ |
|---|---|---|---|
| 50 μm | 7.281e-7 | 1.2977 | 0.999 |
| 80 μm | 1.327e-6 | 1.4590 | 0.992 |
| 110 μm | 1.421e-6 | 1.3872 | 0.993 |
| 220 μm | 2.520e-8 | 2.1137 | 0.999 |

In conclusion, we observe a trade-off between responsivity and spectral flatness. While broadband PVDF membranes for scientific research may prefer a flat spectral response for the purpose of reporting the relative amplitude of spectral components faithfully with minimal deconvolution, industrial applications may find the trade-off for increased sensitivity acceptable for a variety of applications such as therapeutic treatment monitoring and alignment.

## IV. References

[1] Y. Xin, H. Sun, H. Tian, C. Guo, X. Li, S. Wang, C. Wang, "The use of polyvinylidene fluoride (PVDF) films as sensors for vibration measurement: A brief review," Ferroelectrics, vol. 502, 2016, p 28-42.

[2] L. Seminara, M. Capurro, P. Cirillo, G. Cannata, M. Valle "Electromechanical characterization of piezoelectric PVDF polymer films for tactile sensors in robotics applications" Sensors and Actuators A: Physical, Volume 169, Issue 1, 10 September 2011, Pages 49-58.

[3] W.-S. Chen, T. Matula, A. Brayman, and L. Crum "A comparison of the fragmentation thresholds and inertial cavitation doses of different ultrasound contrast agents" J. Acoust. Soc. Am. 113 (1), January 2003

[4] P.C. Beard; A.M. Hurrell; T.N. Mills "Characterization of a polymer film optical fiber hydrophone for use in the range 1 to 20 MHz: A comparison with PVDF needle and membrane hydrophones" IEEE Transactions on Ultrasonics, Ferroelectrics, and Frequency Control, Vol 47. Jan. 2000 p. 256 – 264.

[5] V. Wilkens and W. Molkenstruck "Broadband PVDF Membrane Hydrophone for Comparisons of Hydrophone Calibration Methods up to 140 MHz" IEEE Transactions on ultrasonics, ferroelectrics, and Frequency Control, vol. 54, no. 9, September 2007.

[6] O. A. Sapozhnikov, Y. A. Pishchalnikov, A. D. Maxwell, M. R. Bailey "Calibration of PVDF Hydrophones Using a Broad-Focus Electromagnetic Lithotripter" 2007 IEEE Ultrasonics Symposium p. 112-115.

[7] https://www.acoustics.co.uk/product/radiation-force-balance/

[8] https://www.acoustics.co.uk/pal/wp-content/uploads/2016/05/f28-tds-2019.pdf

[9] R. Preston (Ed.) "Output Measurements for Medical Ultrasound," Springer-Verlag 1991

[10] G. R. Harris, R. C. Preston, and A. S. DeReggi, "The impact of piezoelectric PVDF on medical ultrasound exposure measurements, standards, and regulations," IEEE Trans. UFFC, vol. 47, pp. 1321-1335, 2000.